\documentclass[prb, amsmath, amssymb, twocolumn]{revtex4}
\pdfoutput=1

\usepackage{graphicx}
\usepackage{subfig}
\usepackage{color}
\usepackage{hyperref} 

\hypersetup{
pdfauthor={Rolf W. Reinthaler, Ewelina M. Hankiewicz},
pdftitle={Interplay of bulk and edge states in transport of two-dimensional topological insulators},
pdfsubject={Publication in theoretical condensed matter physics},
pdfproducer={}
}

\newcommand{\ket}[1]{\left\vert #1 \right\rangle}
\newcommand{\bra}[1]{\left\langle #1 \right\vert}

\newcommand{\imag}{\mathrm{i}}
\newcommand{\ek}{\epsilon(k)}
\newcommand{\Mk}{\mathcal{M} (k)}
\newcommand{\UnitM}{\mathbb{I}}
\newcommand{\rd}{\mathrm{d}}
\newcommand{\unit}[1]{\;\mathrm{#1}}
\newcommand{\keff}{k^{\mathrm{eff}}_x }
\newcommand{\akeff}{\langle k ^{\mathrm{eff}}_x\rangle}

\renewcommand{\Im}{\mathrm{Im}}
\renewcommand{\Re}{\mathrm{Re}}
\renewcommand{\vec}[1]{\boldsymbol{#1}}

\begin{document}
\date{\today}

\title{Interplay of bulk and edge states in transport of two-dimensional topological insulators}
\author{R. W. Reinthaler and E. M. Hankiewicz}
\affiliation{
Faculty of Physics and Astrophysics, University of W\"urzburg, W\"urzburg, Germany}


\begin{abstract}
We study transport in two-terminal metal/quantum spin-Hall insulator (QSHI)/metal junctions.
We show that the conductance signals originating from the bulk and the edge contributions are 
not additive. While for a  long junction the transport is determined by the edge states contribution, 
for a short junction, the conductance signal is built from both  bulk and edge states in the ratio
which depends on the width of the sample. Further, in the topological insulator regime the conductance for short junctions
shows a non-monotonic behavior as a function of the sample length. 
Surprisingly this non-monotonic behavior of conductance can be traced to
the formation of an effectively propagating solution which is robust against scalar disorder. 
Our predictions should be experimentally verifiable in HgTe QWs and Bi$_2$Se$_3$ thin films.

\end{abstract}
\maketitle

\section{introduction}
\label{sect:intro}
Topological  states of matter are  characterized by bulk invariants:  Chern numbers \cite{Thouless1982, Kohmoto1985} or by a $Z_2$ invariant \cite{Kane2005b}, depending if the time reversal symmetry is broken or conserved.
 Due to the bulk-boundary correspondence \cite{Halperin1982, Hasan2010} this classification can be translated to the existence of topologically protected states at the edges of the system.
 In the case of the $Z_2$ insulator the edge states are formed by time reversed modes, so called Kramers'  partners and they are helical \cite{Kane2005}, 
i.e. Kramers' partners counter-propagate along a given edge of the sample.  Depending if the number of Kramers' pairs on the edge is even or odd, the system is topologically trivial or non-trivial.  
Although the topological invariant is the property of the bulk,  experiments usually detect edge states properties.
Indeed the confirmation that HgTe/CdTe quantum wells (QWs) are two-dimensional topological insulators \cite{Bernevig06} was provided through the measurement  
of quantized edge conductance \cite{Koenig2007Science, Roth2009, Bruene2011, Koenig2008}.\\

One benefit of HgTe/CdTe QWs is that the topological order can be controlled by the thickness of the HgTe layer: below the critical thickness of $d_c \approx 6.3\unit{nm}$ the system is a trivial insulator,
 whereas above $d_c$ the system behaves as topological insulator. This gives the possibility to observe the topological phase transition directly.
Therefore it is of great interest to find further experimental measurable indicators which distinguish the two regimes. 
Since the topological invariant is a bulk property, it seems natural  that the bulk conductivity could also carry information about topological properties of the system.
Indeed it was proposed in Ref.~\onlinecite{Novik10} that a non-monotonic conductance as a function of geometrical aspect ratio in metal/HgTe/metal junctions with a characteristic maximum describes the topological insulator regime. 
The conductance maximum occurs when incoming metallic bulk states tunnel through a short and wide two-dimensional topological insulator 
and is robust against scalar disorder on the order of the energy gap \cite{Recher10}.
 However, so far this analysis was limited to the periodic boundary conditions (PBC) and neglected the existence of edge states, which can significantly modify the behavior in experimentally relevant setups.
 Here we analyze carefully the properties of the single QSHI/metal interface as well as the interplay between edge states and bulk states in metal/QSHI/metal junctions. 
To do so we apply a generalized wave matching method based on Fourier modes, like it was e.g. used to analyze two interface tunneling junctions in HgTe \cite{Zhang10}.
 Zhang et al. restricted themselves to a low energetic behavior around the gap, which is dominated by the linear dispersion of the bulk and the edge states. 
In contrast our studies combine a topological insulator  with two highly doped metallic leads treated as highly doped HgTe with quadratic dispersion.
We show that the conductance signals originating from the bulk and the edge contributions are 
not additive. While for a  long junction the transport is determined by the edge states contribution, 
for a short junction, the conductance signal is built from both  bulk and edge states in the ratio
which depends on the width of the sample. Further, the conductance for short junctions
shows a non-monotonic behavior as a function of the sample length in the topological insulator regime. 
Surprisingly this non-monotonic behavior of conductance can be traced to
the formation of an effectively propagating solution which is robust against scalar disorder. 
Our predictions should be experimentally verifiable in HgTe QWs and Bi$_2$Se$_3$ thin films.\\
The rest of this paper is organized as follows: In the next Section we give a short introduction to the model. In Section \ref{sect:1int} we analyze the single QSHI/metal interface, which helps us to understand the metal/QSHI/metal junction results in Section \ref{sect:2int}. We use tight-binding calculations in Section \ref{sect:disorder}
to test the robustness of conductance maximum in the presence of scalar disorder. We finish the paper with conclusions.

\section{Brief description of model}
\label{sect:model}
Close to the $\Gamma$ point ($k=0$), HgTe quantum wells or Bi$_2$Se$_3$ thin films can be described by an effective $4\times4$ Dirac-like model \cite{Bernevig06, Rothe10, Liu2010a}. In the following we concentrate on the description of HgTe QWs while we will comment on Bi$_2$Se$_3$ thin films at the end of this section.
In the absence of structure inversion asymmetry the Hamiltonian for HgTe QWs 
consists of two decoupled blocks, which are related by time reversal symmetry (TRS). 
Using the basis $(\ket{E1+}, \ket{H1+}, \ket{E1-}, \ket{H1-})^T$, with (E1) and (H1)  designating electron and heavy hole sub-bands and $\pm$ denoting the Kramers' partner, the Hamiltonian reads ($k_\pm = k_x \pm \imag k_y$, $k=\sqrt{k_x^2 +k_y^2}$, $k_\parallel = (k_x,k_y)^T$)
\begin{align}
	\label{eq:Hamiltonian}
	\mathcal H = \begin{pmatrix}
			h(k_\parallel) & 0 \\
			0 & h^*(-k_\parallel) \\
		\end{pmatrix}, \quad h(k_\parallel) = \epsilon (k) \UnitM + d_{\alpha} (k_\parallel) \sigma_{\alpha},
\end{align}
with $d_\alpha (k_\parallel) = (Ak_x, -Ak_y, \Mk)^T$. $\sigma_\alpha$ are the Pauli matrices and $\UnitM$ is the unit matrix in the subband space.
Here $\ek=C-D k^2 $ is the energy dispersion, $\Mk = M - B k^2$ is the $k$ dependent gap and $A$ gives the strength of the coupling between electron and heavy hole states. The parameters $A$, $B$, $D$ and $M$ depend on the quantum well width. Below (above) the critical thickness of the quantum well  $M > 0$ ($M<0$), the system describes the topologically trivial (non-trivial) regime. Due to the block diagonal form of Eq.~\eqref{eq:Hamiltonian} we will just consider the upper block in the following. The results for the lower block then follow by TRS. A generalization including a finite value of structure inversion asymmetry was presented in Ref.~\onlinecite{Rothe10}.\\

In the following we consider the transport along $x$-direction through a single interface between QSHI and normal metal as well as  QSHI connected to two metallic  leads 
(see inset to Fig.~\ref{fig:1IntCond} and Fig.~\ref{fig:2IntSetup}, respectively).
We set C to zero in the center region, so called quantum spin-Hall insulator regime, and to a large negative value in the leads, thereby
effectively modeling the metal contacts by highly doped HgTe, cf. Fig.~\ref{fig:2IntSetup} for a schematic. 
So far similar systems were described within periodic  boundary conditions (PBC) \cite{Novik10} where
due to the periodicity in y-direction $k_y$ is conserved and quantized i.e. $k_y = 2 \pi n/W$, where $n = 0,\pm 1 , \ldots , \pm N_{\mathrm{max}}$ with $N_{\mathrm{max}}$ given by the number of propagating modes.
 However PBC do not allow for a formation of edge states and the competition between edge and bulk transport was not explored so far. 
Therefore in this paper, in order to include edge states we choose hard wall boundary conditions (HBC) i.e. the wave function is zero at the edges of the sample.
 An analytical solution of Eq.~\eqref{eq:Hamiltonian} for HBC \cite{Zhou08} couples different $k_y$ modes and a simple wave matching procedure like e.g. in Refs.~\onlinecite{Yokoyama09} and \onlinecite{Novik10} is no longer possible.
 To resolve this issue we expand the states in terms of the Fourier modes $\phi_n(y) = \sqrt{2/W} \sin\left[n \pi y/W \right]$, whose orthogonality allow an independent matching for each mode.
 Therefore, for an infinite quasi-1D system with $k_x$  conserved, the eigenvector can be expanded as follows:
\begin{align}
	\label{eq:FourierAnsatz}
	\psi_m (x,y)=  \exp\left[\imag k_x^m x  \right] \sum_{n  = 1}^\infty \chi_{n}^m \phi_n(y)
\end{align}
Here the index $m$ labels the $k_x^m$-eigenvalues, $\chi_n^m$ is the two component spinor corresponding to the upper block of Hamiltonian (1) for a given transverse mode $n$. To find the eigenvalues and eigenspinors we use the Schr\"odinger equation 
\begin{align}
	(\mathcal H - E)  \exp\left[\imag k_x^m x  \right] \sum_{n_1  = 1}^\infty \chi_{n_1}^m \phi_{n_1}(y) = 0.
\end{align} 
Multiplying from the left by $\int_0^W \rd y \phi^\dagger_{n_2}(y)$ and using the orthogonality of the sine functions yields:
\begin{align}
\label{eq:EigenvalueProblemNoMat}
  \mathcal{H}^{\text{const}} \chi^m_{n_2} &+ \mathcal{H}^{k_x} k^m_x \chi^m_{n_2}  + \mathcal{H}^{k_x^2} k^m_x \chi^{\prime m}_{n_2} \nonumber \\
  &+ \sum_{n_1}\int_0^W \rd y\phi_{n_2}^\dagger \mathcal{H}^{k_y} \phi_{n_1} \chi^m_{n_1} = 0 ,
\end{align}
where all terms of the Hamiltonian proportional to $k_y = -\imag \partial_y$ are collected in $\mathcal{H}^{k_y}$, all constant terms in $ \mathcal{H}^{\text{const}}$ and all with $k_x$ ($k_x^2$) in $ \mathcal{H}^{k_x}$ ($\mathcal{H}^{k_x^2}$). Additionally we introduced $\chi_{n}^{\prime m} = k_x^m \chi_n^m$. Defining the vectors $\chi^m = (\chi^m_{n=1}, \chi^m_{n=2}, \ldots)^T$ and $\chi^{\prime m} = (\chi^{\prime m}_{n=1}, \chi^{\prime m}_{n=2}, \ldots)^T$, which are  built by the $2$ component spinors $\chi_n^m$ and $\chi_n^{\prime m}$, Eq.~\eqref{eq:EigenvalueProblemNoMat} can be written as matrix equation
\begin{align}
	\label{eq:EigenvalueProblem}
	\begin{pmatrix}
		\UnitM & 0\\ 0 & \left(H^{k_x^2}\right)^{-1}\\
	\end{pmatrix}
	\begin{pmatrix}	
		0 &  \UnitM \\ H^{\text{const}} + H^{k_y} & H^{k_x} \\
	\end{pmatrix} 
	\begin{pmatrix} \chi^m \\ \chi^{\prime m} \end{pmatrix}
		=
	k_x^m 	
	\begin{pmatrix} \chi^m \\ \chi^{\prime m} \end{pmatrix}.
\end{align}
So doubling the dimension of the system of equations by introducing $\chi^{\prime m}$ allows to reduce the problem of finding $k_x^m$ to a linear eigenvalue equation\cite{Chang1982, Liu2011}. The sub-matrices in Eq.~\eqref{eq:EigenvalueProblem} are 
\begin{align}
	H^{k_x^2}_{n_1n_2} &= \delta_{n_1n_2} 
	\mathrm{diag} \left[\tilde{D}_+,\tilde{D}_- \right],\\
    H^{k_x}_{n_1n_2} &=\delta_{n_1n_2}
    \begin{pmatrix}
		0 &A\\
		A &0\\
	\end{pmatrix}, \\
	H^{\text{const}}_{n_1n_2} &=\delta_{n_1n_2} \mathrm{diag} \left[\tilde{E}_+,\tilde{E}_- \right],\\
	H^{k_y}_{n_1n_2} &= 
	\begin{pmatrix}
		-\left(\frac{n \pi }{W}\right)^2\tilde{D}_+ \delta_{n_1n_2} & \imag A \eta_{n_1n_2} \\
		-\imag A \eta_{n_1n_2} & -\left(\frac{n \pi }{W}\right)^2\tilde{D}_- \delta_{n_1n_2} \\
    \end{pmatrix},
\end{align}
Further we used here $\tilde{D}_\pm = (D \pm B)$ and $\tilde{E}_\pm = C -E \pm M$. The only term coupling different modes is $\eta_{n_1n_2} = \bra{\phi_{n_1}} k_y \ket{\phi_{n_2}}$. \\
The solutions $k_x^m$ can be characterized as propagating ($k_x^m \in \mathbb{R}$) or evanescent ($\Im(k_x) \neq 0$).
For real $k_x^m$ the propagation direction can be determined by the sign of 
\begin{align}
v^m = \int_0^W \rd y \psi_m^\dagger (x,y)\left[\partial_{k_x} h(k)\right]_{k_x \rightarrow k_x^m}  \psi_m (x,y).
\end{align} 
Analogously evanescent states with $\Im(k_x)>0$ ($\Im(k_x)<0$) are decaying to the right (left). 
In the following we will denote right (left) going solutions by the index $m_R$ ($m_L$). In general the Fourier ansatz also works when the two blocks of Hamiltonian \eqref{eq:Hamiltonian} are connected by the Rashba-spin orbit coupling  like it was e.g. done in Ref.~\onlinecite{Zhang10}.\\
Having the solutions for a single system we can now consider QSHI/metal and metal/QSHI/metal junctions. 
The interfaces in Figs.~\ref{fig:1IntCond} and \ref{fig:2IntSetup} break translational symmetry in $x$-direction and an incoming state $\psi_{\text{in}}$ can be scattered in all possible $k_x$-modes. We will now label the $k_x$ eigenvalues as:
$k_{x,1}^m$ for the part 1 stretching from $x = - \infty$ to $x = - L/2$, $k_{x,2}^m$  for $-L/2<x<L/2$ and $k_{x,3}^m$  for $x \ge L/2$. If the incoming state is assumed as incident from the left in mode $m$, the state in the left lead takes the form
\begin{align}	
	\label{eq:LeftLeadState}
	\Psi_1 (x,y) = \psi_{m,1}(x,y) + \sum_{m_L} r_{m_L, m} \psi_{m_L,1}(x,y)
\end{align}
and that of the right lead is
\begin{align}
	\label{eq:RightLeadState}
	\Psi_3(x,y) = \sum_{m_R} t_{m_R,m} \psi_{m_R,3}(x,y).
\end{align}
Here $\vert r_{m_L, m}\vert^2 $ and $\vert t_{m_R,m}\vert^2$ are the probabilities that the incoming mode $m$ is reflected into mode $m_L$ of the left lead or transmitted into mode $m_R$ of the right lead, respectively. 
Only in the central region of length $L$ there are left and right outgoing states
\begin{align}
	\label{eq:SampleState}
	\Psi_2(x,y) = \sum_{m_R}c_{m_R,m} \psi_{m_R,2} (x,y) +\sum_{m_L} d_{m_L, m} \psi_{m_L,2}(x,y).
\end{align}
The scattering amplitudes are computed by matching the wave functions $\Psi_i(x,y)$ and the associated currents $J^x_i (x,y)= \left[\partial_{k_x} h(k)\right]_{k_x \rightarrow -\imag \partial_x} \Psi_i(x,y)$ at the interfaces, i.e. $\Psi_1(-L/2, y) = \Psi_2(-L/2, y)$, $\Psi_2(L/2, y) = \Psi_3(L/2, y)$, $J^x_1(-L/2, y) = J^x_2(-L/2, y)$ and $J^x_2(L/2, y) = J^x_3(L/2, y)$. For setups like in Fig.~\ref{fig:1IntCond}, where we have only one interface ($L=0$), the matching simplifies to $\Psi_1(0, y) = \Psi_3(0, y)$ and $J^x_1(0, y) = J^x_3(0, y)$. The total transmission then is 
\begin{align}
	\label{eq:transmission}
	T = \sum_{\{m_R \left\vert k_{x,3}^{m_R} \in \mathbb{R} \right.\}} \sum_{\{n_R \left\vert k_{x,1}^{n_R} \in \mathbb{R} \right.\}} \frac{v^{m_R}_3}{v^{n_R}_1} \vert t_{m_R, n_R}\vert^2.
\end{align}
Moreover the knowledge of all scattering amplitudes determines the full state $\Psi(x,y)$ up to normalization. $\Psi(x,y)$ is the response to an incoming mode $\psi_{m,1}(x,y)$ from the left lead. Therefore it allows us to calculate the non-equilibrium charge density $n (x,y) = \vert\Psi(x,y)\vert^2$ as well as the non-equilibrium current density, $\vec{J}(x,y)= \Psi^\dagger(x,y) \nabla_{\vec{k}} h(k) \Psi(x,y)$. 
\\
For actual computation  the Fourier series has to be truncated at a sufficiently high mode $N$. The error of this approximation depends weakly on the width of the sample and on the Fermi energy. However, the asymmetric shape of an edge state is harder to approximate by sine functions than bulk states. Nevertheless a good approximation can be achieved. For example the maximal relative errors in the 99 lowest $k_x^m$ eigenvalues at $W= 1000\unit{nm}$ and $E_f = 0$ changes from 2 \% at $N = 100$ to 0.6 \% at $N = 200$. The higher modes decay very fast and have little influence on the transport. Unless otherwise specified we will use the following parameters: $A = 0.375\unit{nmeV}, \; B = -1.120 \unit{nm^2 eV}, \; D = -0.730 \unit{nm^2 eV}, \; M = - 3.10\unit{meV}$. When we are referring to the normal regime we put $M = +3.10\unit{meV}$.\\

In Bi$_2$Se$_3$ thin films the overlap of surface states of opposite surfaces leads to the opening of a gap at $\Gamma$ point\cite{Linder2009, ZhangY2010}. 
This gap closes as the energetically lowest lying states, which have opposite parity, cross at certain thicknesses $d_z^n$ $(n=1,2,\ldots)$ of the thin film  \cite{Liu2010a}. 
At each crossing the parity changes leading to the oscillations between topologically trivial and non-trivial insulators as a function of the film thickness. Around the band crossing points a folding of the Hamiltonian of the 3D topological insulator \cite{Zhang2009, Liu2010b} to the lowest energetic sub-bands reproduces the model Eq.~\eqref{eq:Hamiltonian} \cite{Liu2010a, Lu2010, Shan2010} for the 2D topological insulator. We choose for the calculations $d_z = 3\unit{nm}$ with the parameters of  2D effective model as follows: $A = 0.406\unit{nmeV}, \; B = -0.250 \unit{nm^2 eV}, \; D = 0.070 \unit{nm^2 eV}, \; M = -22.5\unit{meV}$ \cite{Lu2010}.

\section{Single interface between QSHI and metal}
\label{sect:1int}
\begin{figure}
\includegraphics[width=0.9\linewidth]{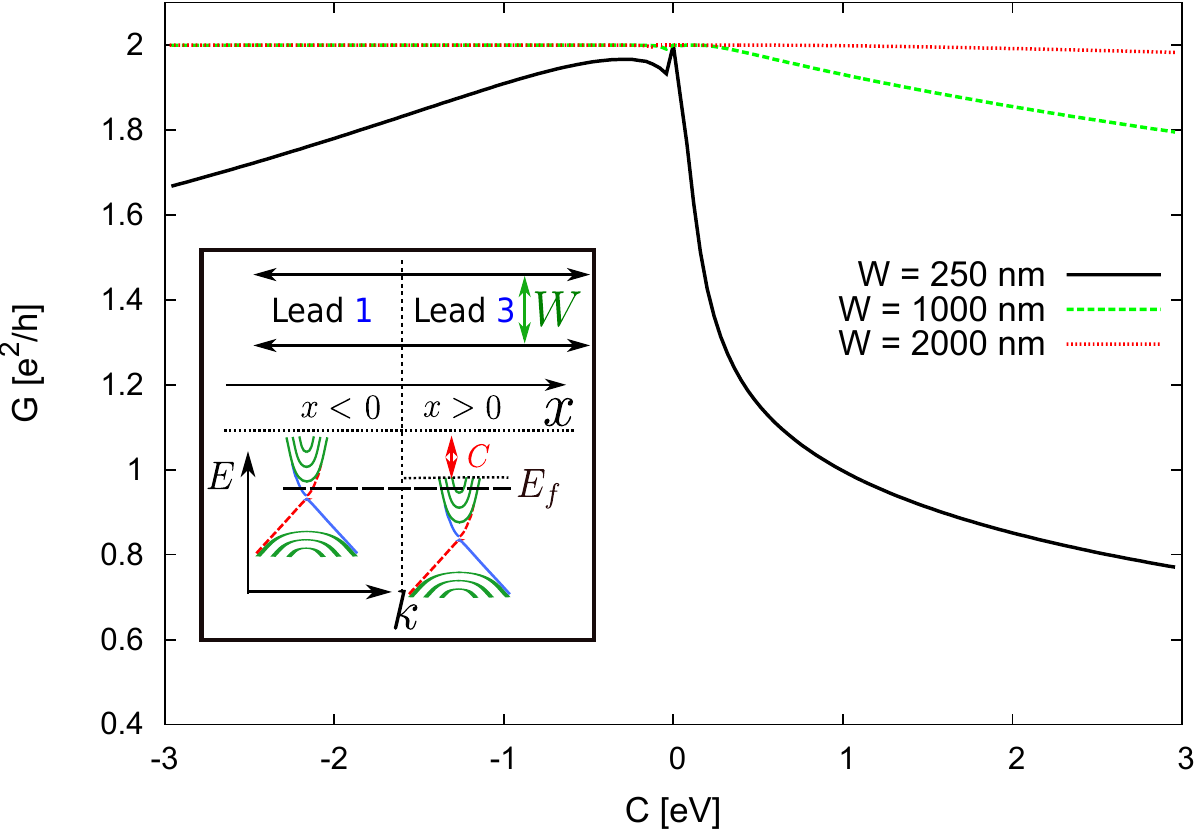}
\caption{	
The conductance as a function of the energetic height of the interface at a junction between a QSHI and a doped QSHI. A sketch of the system is shown in the inset.
 Perfect transmission is achieved in wide samples where counter propagating edge states for the same Kramers' partner (spin) do not couple.
 Negative (positive) $C$ indicate a QSHI/n-type metal (QSHI/p-type metal) interface. }
\label{fig:1IntCond}
\end{figure}

Here we analyze the injection of the helical edge states from the quantum spin Hall insulator into a metallic lead. The latter is modeled by Eq.~\eqref{eq:Hamiltonian} with a high doping, i.e. large $\vert C\vert$. The used setup is sketched in the inset of Fig.~\ref{fig:1IntCond}. On the QSHI side, the Fermi energy is chosen to be zero while on  the metal side $E_f$ is shifted by $C$, with $C$ characterizing an energetic height (strength) of the interface. 
In Fig.~\ref{fig:1IntCond} we show the transmission through such an interface as a function of $C$ for different widths of the system $W = 250, \; 1000$ and $2000 \unit{nm}$. 
While for narrow samples strong interfaces introduce a significant backscattering due to an overlap of counter-propagating edge states, perfect transmission is observed for wide samples.  
The perfect transmission for wide samples can be explained analogously as for the topologically protected edge states in graphene \cite{Prada10}, where it was found that the counter propagating edge states for the same Kramers' partner  are orthogonal.
 Indeed taking into account an exponential decay of the wave function for an edge state \cite{Zhou08} it can be easily confirmed that the overlap between edge states goes to zero for $W\rightarrow \infty$ in the presence of any $x$ dependent potential:
\begin{align}
	\label{eq:overlap}
	\bra{\Psi_{k_x}}V(x)\ket{\Psi_{-k_x}} = V(x)\int_0^W \rd y \Psi^\dagger_{k_x}\Psi_{-k_x} \stackrel{W \rightarrow \infty}{\longrightarrow}0.
\end{align}
\\
By the Onsager- Casimir symmetry relation \cite{Buettiker1986, Buettiker1988} injecting electrons from the metal to QSHI should be also perfect for very wide samples.
Let us mention for completeness, that a linear Rashba coupling \cite{Rothe10} does not increase the backscattering, because, although 
the direction of edge state polarization is rotated, there are still two orthogonal solutions for edge states at the same edge.

\section{Metal/QSHI/metal Junction}
\label{sect:2int}
\begin{figure}
\includegraphics[width=0.9\linewidth]{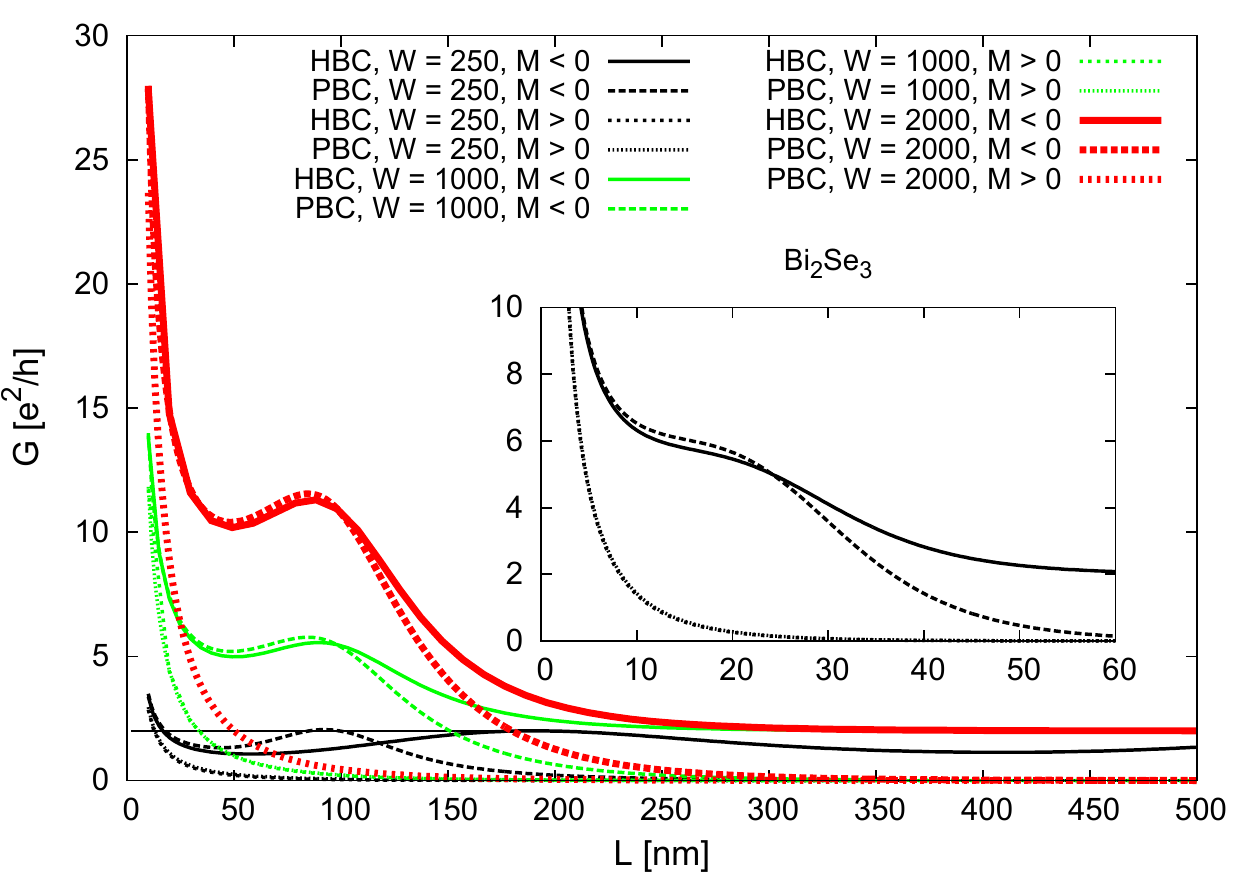}
\caption{	
The conductance through a metal/QSHI/metal junction, like depicted in Fig. \ref{fig:2IntSetup}, as a function of the sample length $L$ for different $W$ (all values in $\unit{nm}$). We compare the normal regime ($M = 3.10 \unit{meV}>0$) with the inverted one ($M = -3.10 \unit{meV}<0$) for HBC and PBC. Metallic leads are treated as highly doped HgTe with $C = -2.5 \unit{eV}$. 
The inset shows the corresponding conductance  signals for a Bi$_2$Se$_3$ thin film ($M = \pm 22.5 \unit{meV}$) of width $250\unit{nm}$ at $C=-2.5 \unit{eV}$. 
}
\label{fig:2IntCond}
\end{figure}

\begin{figure}
 \subfloat[]{\includegraphics[width = 0.45\linewidth]{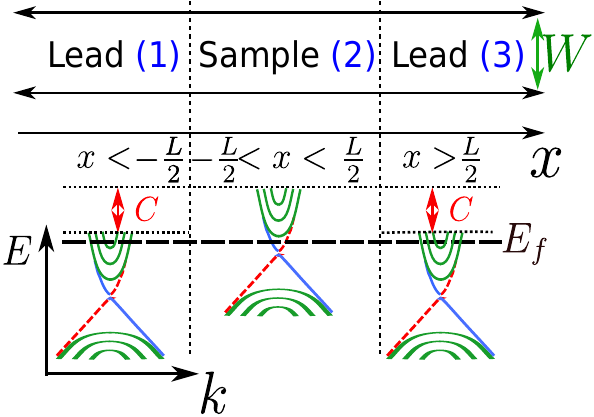}\label{fig:2IntSetup}}\hfill
 \subfloat[]{\includegraphics[width = 0.55\linewidth]{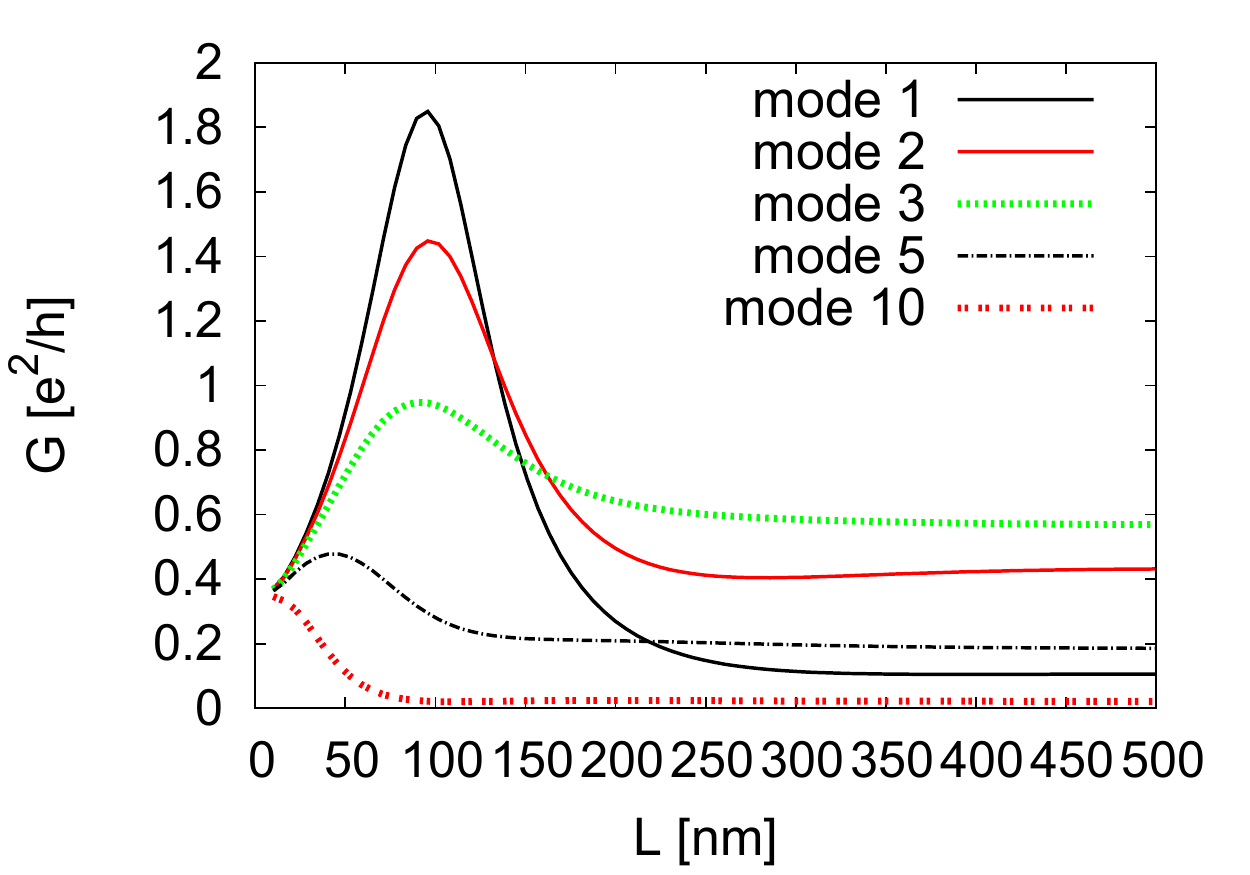}\label{fig:2IntModes}}
 \caption{(a) A schematic of the setup, consisting of the two metallic leads and the central QSHI at zero Fermi energy. The energy shift $C$ is also shown in the band structures of the individual segments. (b) The conductance as a function of $L$ for single modes ($W=1000\unit{nm}$ and $C= -2.5\unit{eV}$). The maximum is mainly carried by the low modes.}
 \label{fig:SetupModes}
\end{figure}

In this section we consider metal/QSHI/metal junctions with a finite length $L$ of the QSHI as shown in Fig.~\ref{fig:2IntSetup}. So far such junctions
 have been only analyzed using PBC which neglect the topological edge states \cite{Novik10}. 
  In particular it was found that such junctions allow for the distinction of different topological phases purely due to evanescent bulk modes \cite{Novik10}. 
While the conductance rises monotonically with decreasing $L$ in the topologically trivial regime ($M>0$) , the topologically non-trivial regime ($M<0$) is characterized by a conductance maximum 
at $L_{\text{max}}$
. In the limit of $D=0$ and $BC \gg A^2 \gg BM$ ($C<0$) the position of the maximum can be predicted for the zero mode by 
\begin{align}
	\label{eq:Lmax}
	L_{\text{max}} \approx \frac{A}{2\vert M\vert} \ln\left(\frac{2BC}{A^2} \right).
\end{align}
 The purpose of this paper is to better understand the existence of the conductance maximum in the topologically non-trivial regime as well as to study the interplay of the bulk and edge contributions using hard wall  boundary conditions.
 We use the technique described in Ref.~\onlinecite{Novik10} to solve the problem with PBC and  the method described in Section \ref{sect:model} for a full solution with the hard wall boundary conditions. 
In Fig.~\ref{fig:2IntCond} we present the conductance as a function of the sample length $L$. 
First we observe that the conductance decreases exponentially  with the increase of $L$ in the normal regime, independent of the boundary conditions. Indeed in this regime we expect only evanescent solutions and therefore the same results for PBC and HBC.
We now focus on the topological insulator regime i.e.  inverted regime with $M<0$. 
In this regime, for HBC, both evanescent and  edge states contributions to the conductance are present and the signal depends
on the length and the width of the sample.

\begin{figure}
 \subfloat[]{\includegraphics[width = 0.375\linewidth]{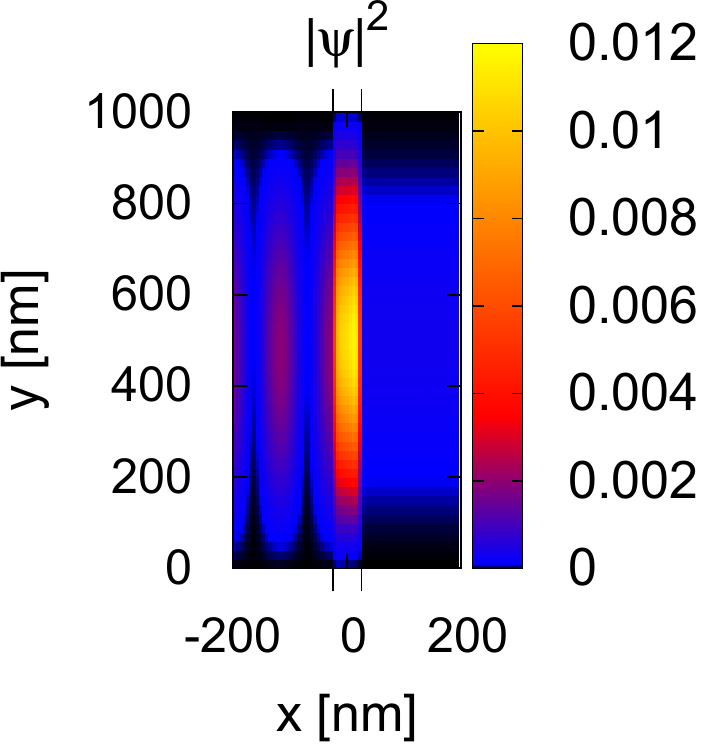}\label{fig:2IntDen_a}}\hfill
 \subfloat[]{\includegraphics[width = 0.375\linewidth]{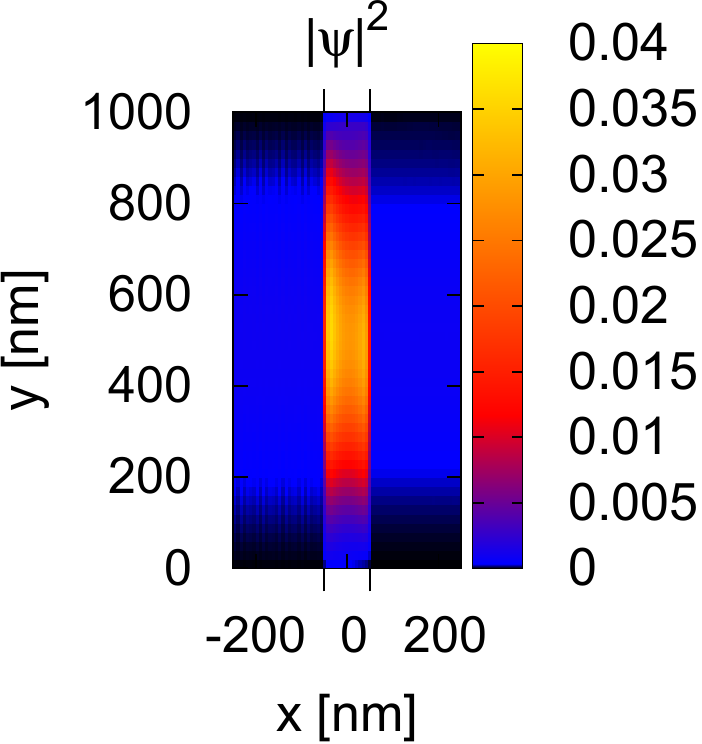}\label{fig:2IntDen_b}}\\
 \subfloat[]{\includegraphics[width = 0.485\linewidth]{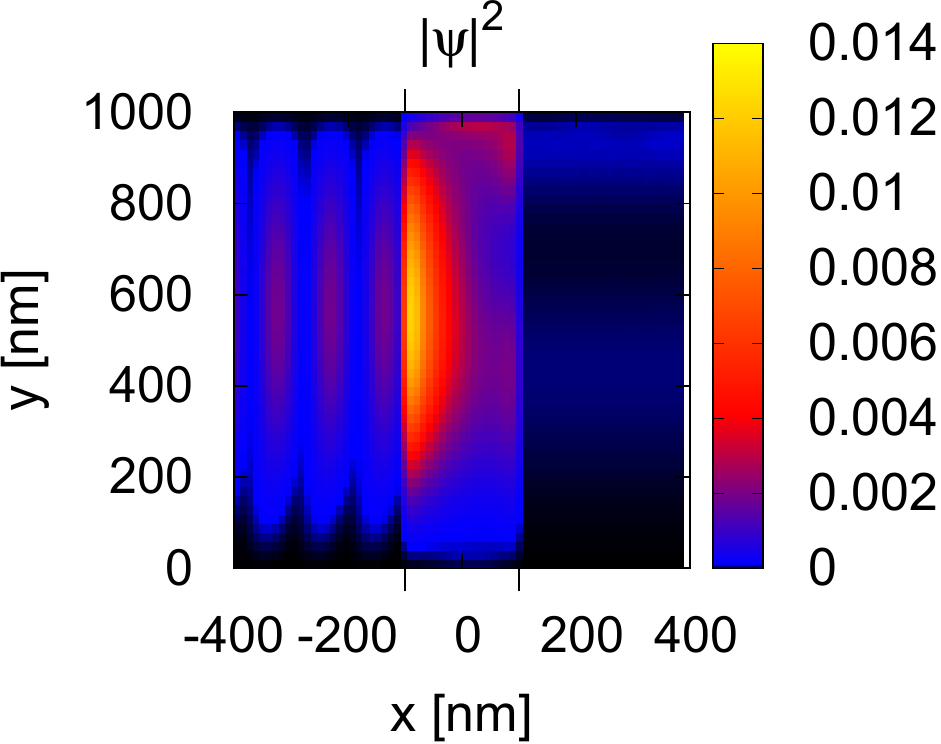}\label{fig:2IntDen_c}}\hfill
 \subfloat[]{\includegraphics[width = 0.485\linewidth]{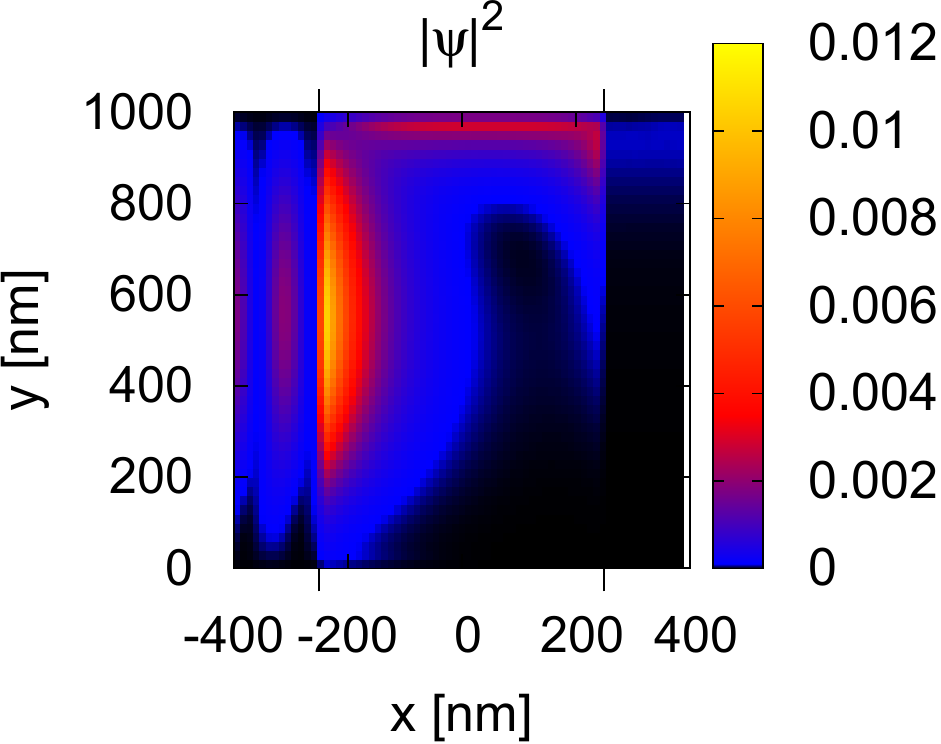}\label{fig:2IntDen_d}}\\
\caption{	
Evolution of charge density in a metal/QSHI/metal junction as a function of the length of the sample. The charge response is the response to the first incoming mode, cf. Sec.~\ref{sect:model} after Eq.~\eqref{eq:transmission}. The interfaces at $\pm L/2$ are indicated by two black lines. In the plot, only upper edge state is seen since we restricted our calculations to the one spin direction. We used $W=1000\unit{nm}$ and $C= -2.5\unit{eV}$. (a) $L=50\unit{nm}$, (b) $L=100\unit{nm}\approx L_{\text{max}}$, (c) $L=200\unit{nm}$ and (d) $L=500\unit{nm}$.}
\label{fig:2IntDen}
\end{figure}

\begin{figure}
\includegraphics[width=0.9\linewidth]{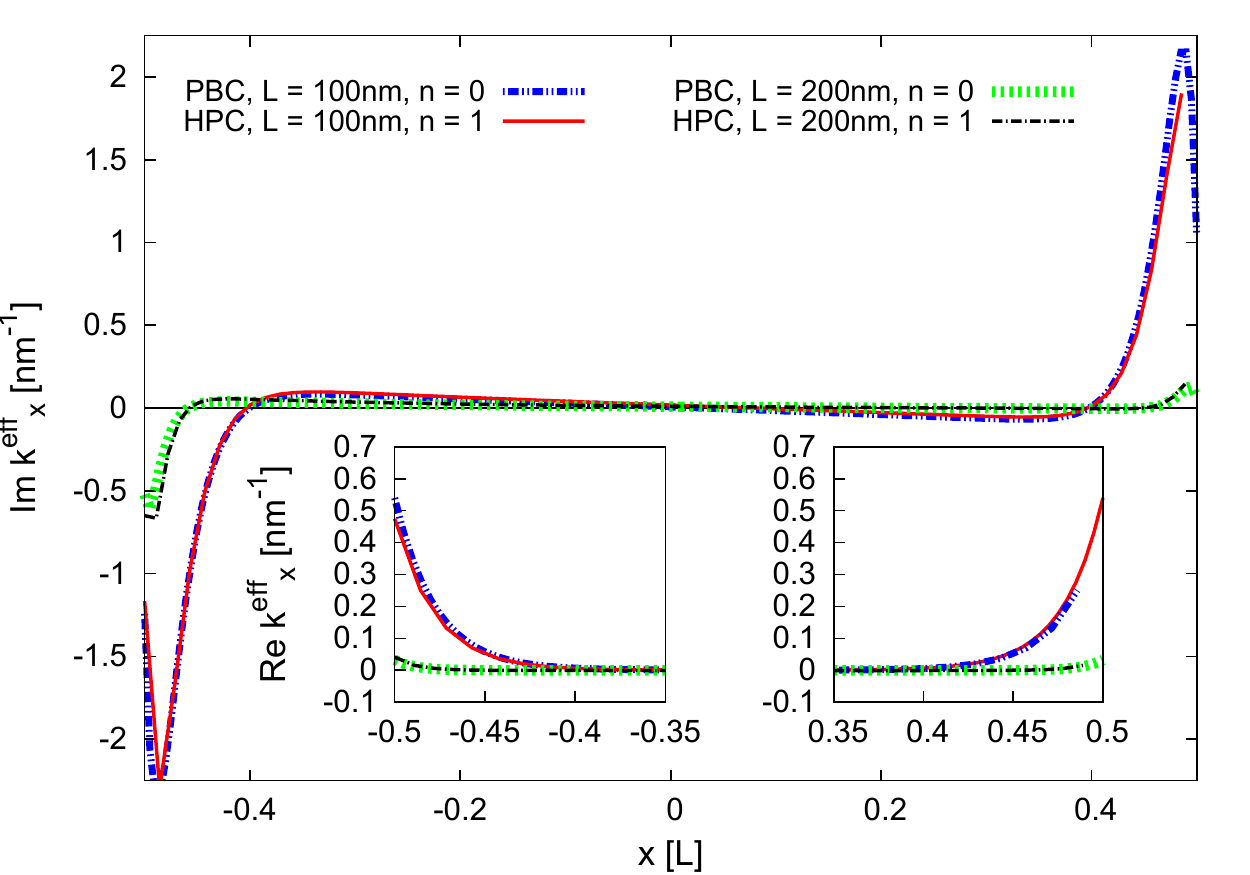}
\caption{The imaginary and real (inset) part of $\keff$ value as a function of the position $x$ for different lengths of QSHI part, boundary conditions and incoming modes $n$. We used $W=1000\unit{nm}$ and $C= -2.5\unit{eV}$. The conductance maximum corresponds to the length $L_{\text{max}}\approx 100\unit{nm}$.}
\label{fig:keff}
\end{figure}

\begin{figure}
\includegraphics[width=0.9\linewidth]{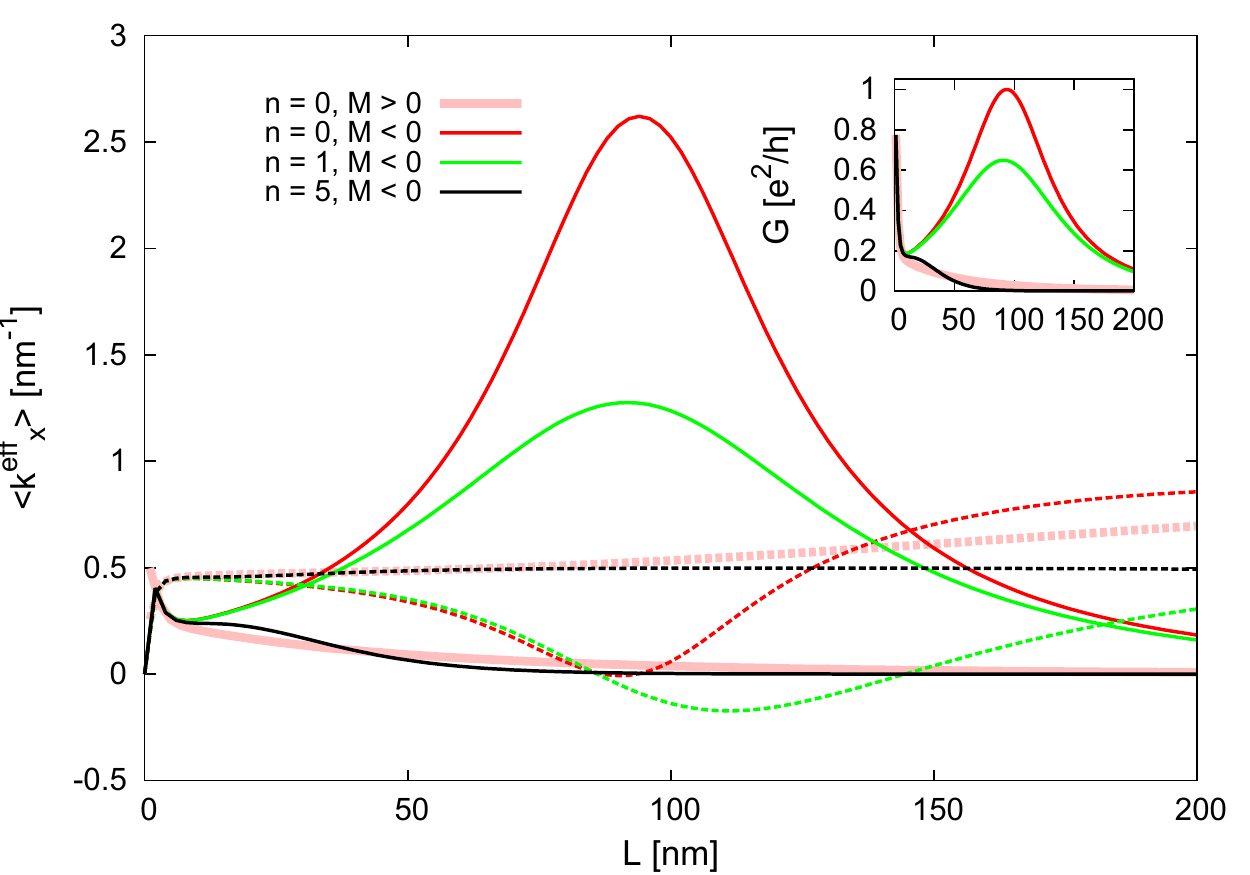}
\caption{$\akeff$ as a function of the length for different incoming modes. Solid and dashed  lines show real and imaginary parts of $\akeff$,  respectively. 
For clarity, only the calculations for PBC are shown. We see that for low modes the position of the conductance  maximum (see inset) coincides with a maximum in the real part and a zero of the imaginary part of $\akeff$. For higher modes this behavior gradually disappears. We used $W=1000\unit{nm}$ and $C= -2.5\unit{eV}$. The maximum lies at $L_{\text{max}}\approx 100\unit{nm}$.}
\label{fig:akeff}
\end{figure}

In the very narrow QSHI regime, see  black line in Fig.~\ref{fig:2IntCond}, we find Fabry-Perot like oscillations in the transmission of the edge states for HBC. At large $L$, the differences between two oscillation maxima is given by $\pi/k_{\text{edge}}$, where $k_{\text{edge}}$ is real wave vector of the edge states. The Fabry-Perot oscillations originates from the backscattering of edge states
at the interfaces and therefore are only present for narrow samples with the finite overlap between counter propagating edge states.  In this case the existence of edge states has a dramatic effect on
the behavior of conductance as a function of L. 
The quantum confinement in narrow wires shifts the energy levels of the states to higher values. This also increases the corresponding imaginary $k_x$ values, which leads to a faster decay of incoming electrons in the QSHI tunneling junction. The combination of the Fabry Perot oscillations of the edge states and the bulk modes changes the position and the shape of the conductance maximum, which does not coincide with the maximum generated by either the evanescent bulk states or the Fabry Perot oscillations alone.  
Therefore 
Eq.~\eqref{eq:Lmax} can not be used to find a maximum of conductance. \\
In contrast,  for a wide and long QSHI part, the conductance is  dominated by the edge states contribution and equals $2e^2/h$. For wide and medium length of QSHI, PBC and HBC give very different results while for short and wide QSHI the transport 
is dominated by tunneling evanescent modes \footnote{Please note, that the maximum in Fig.~\ref{fig:2IntCond} cannot be predicted by Eq.~\eqref{eq:Lmax} even for PBC, since we included a finite $D$ parameter in our calculations.}.
This is a very interesting result since naively one would expect that the edge state and bulk contributions to the conductance are additive for wide QSHI. \\
The same behavior can be found for Bi$_2$Se$_3$ thin films. In the inset of Fig.~\ref{fig:2IntCond} we show the conductance for a $250\unit{nm}$ wide strip. Instead of a maximum we observe a plateau like behavior in the inverted regime. However, increasing further the $C$ parameter restores the conductance maximum. Despite the narrow width the signal of HBC and PBC coincides around the plateau. This can be explained by the larger gap in Bi$_2$Se$_3$, which decreases the overlap of the edge states \cite{Zhou08}.\\
 Since we showed in the last section that the transmission through a sufficiently wide metal-QSHI interface is perfectly quantized, the scattering from the interfaces cannot be the reason for the absence of the edge state contribution. Therefore to understand the non-additive behavior of bulk and edge contributions and the formation of the conductance maximum, we plot in Fig.~\ref{fig:2IntDen} the evolution of the charge density for different sample lengths.
In Fig.~\ref{fig:2IntDen}, we take into account only scattering of the first incoming mode with n=1, like it was described in Sec.~\ref{sect:model} after Eq.~\eqref{eq:transmission}, and use again the parameters for HgTe QWs.
 Since the lowest modes give the largest contribution to the conductance maximum as shown in Fig.~\ref{fig:2IntModes}, it  is justified to analyze first incoming mode signal to grasp the important details of the conductance maximum. For plotting the charge density we use the following normalization:
 \begin{align}
 	\int_0^W \rd y \Psi^\dagger (x = -\frac{L}{2}, y)\Psi (x = -\frac{L}{2}, y) = 1.
 \end{align}
Let us first concentrate on the Figs.~\ref{fig:2IntDen_a}, \ref{fig:2IntDen_c} and \ref{fig:2IntDen_d}, where $L \neq L_{\text{max}}$.
Due to the large $C$ parameter a strong reflection occurs at the left interface leading to standing waves in the left lead, see e.g. Fig.~\ref{fig:2IntDen_d}. 
Further we observe that the density is peaked inside the sample. The incoming states from the left lead are purely metallic bulk ones since for high energies (high C parameter) there is no influence of the edge states.
 In the short QSHI part (see Fig.~\ref{fig:2IntDen_a}), these metallic states are not able to adjust to the shape of edge states and so they keep their form and tunnel through the sample. Therefore for short L the conductance signals from HBC and PBC should coincide as indeed found in Fig.~\ref{fig:2IntCond}  . When the length of QSHI part is comparable to the decay length of the evanescent modes, (see Fig.~\ref{fig:2IntDen_c}), electrons are more likely transmitted through the sample by scattering from the bulk states into the edge states. 
For longer QSHI parts (see Fig.~\ref{fig:2IntDen_d}) transport is driven by edge states.\\
It is interesting to compare Figs.~\ref{fig:2IntDen_a}, \ref{fig:2IntDen_c} and \ref{fig:2IntDen_d} with Fig.~\ref{fig:2IntDen_b}, where we choose the length of QSHI part corresponding to the peak in conductance i.e. $L\approx L_{\text{max}}$.
One can  see that the charge density is much larger in the QSHI part in comparison with other cases where $L \neq L_{\text{max}}$. 
Additionally we do not observe standing waves in the left lead, which indicates  a weaker interface reflection. 
It also becomes clear that at the position of the maximum the edge states are not yet populated, which is the reason for a coincidence of  conductance signals for PBC and HBC.\\
To further investigate the origin of the conductance maximum, when bulk states dominate transport, we analyze the effective wave vector in the direction of propagation for the QSHI part
\begin{align}
	\label{eq:keff}
	\keff(x) = \int\limits_0^W \rd y \Psi_2^\dagger(x,y) \left(-\imag \partial_x\right) \Psi_2(x,y),
\end{align}
where $\Psi_2(x,y)$ is the full scattering solution from Eq.~\eqref{eq:SampleState}. Our analysis will focus on the lowest modes in PBC and HBC, which give the largest contribution to the signal.
In Fig.~\ref{fig:keff} we present the imaginary as well as the real (inset) part of $\keff (x)$ for different lengths and  the lowest incoming mode. 
The lowest modes for PBC and HBC are n=0 and n=1, respectively. In general $\keff$ can exhibit a finite real part even for PBC, i.e. in absence of edge states, due to evanescent solutions combined with complex amplitudes in $\Psi_2(x,y)$. 
Indeed one can see in Fig.~\ref{fig:keff} that the real part of  $\keff$ is non-zero and decays exponentially away from the two interfaces.  The imaginary part instead has a more sophisticated behavior with several crossings of zero. In particular $\Im \keff(x)$ shows an antisymmetric behavior in respect to the sample middle for the length corresponding to the conductance maximum i.e. $L=100\unit{nm} \approx L_{\text{max}}$ for lowest modes in  PBC and HBC designated by blue/dark gray dashed and red/gray solid lines, respectively. 
This suggests that the imaginary part of $\keff$ vanishes after averaging over x-direction. We define  $\akeff$ as:
\begin{align}
	\label{eq:akeff}
	\akeff = \int\limits_{-L/2}^{L/2} \rd x \keff(x)
\end{align}
Fig.~\ref{fig:akeff} shows $\akeff$ as a function of the insulator length for PBC. The dashed red/dark gray curve in Fig.~\ref{fig:akeff} shows that $\Im \akeff$ 
vanishes indeed at $L_{\text{max}}$ for the zero mode. The vanishing evanescent part of the effective wave vector coincides  with a maximum of  the real part of $\akeff$ (solid red/dark gray line),
 which gives the impression of an effectively propagating state leading to the conductance maximum. Away from $L_{\text{max}}$, $\Re \akeff$ decreases while  $\Im \akeff$  has a finite value for the zero mode. 
Indeed Fig.~\ref{fig:keff} shows that the signal  for $\Im \keff$  and   $L= 2*L_{\text{max}}$ is not antisymmetric.  
The inset to Fig.~\ref{fig:akeff} shows that the higher modes contribute less to the conductance maximum, but for $n=1$ there is still a $\Re \akeff$  maximum around $L_{\text{max}}$ (solid green/light gray line in Fig.~\ref{fig:akeff}). 
This maximum for $n=1$ is smaller in comparison to $n=0$ and  is shifted the same way as respective conductance maxima for $n=0$ and $n=1$ (see inset to Fig.~\ref{fig:akeff}) . 
$\Im \akeff$ for $n=1$  has also a zero value at $L_{\text{max}}$.  In contrary for $n=5$, which does not contribute to the maximum, the $\Im \akeff$  (black dashed line) is constant and the $\Re \akeff$ (black solid line) drops to zero. 
Therefore higher modes ( modes with $n>4$ for our set of parameters) behave almost like modes in the topologically-trivial regime (compare  rose/light gray thick lines with black lines in Fig.~\ref{fig:akeff}) and have evanescent character. 
\\

\begin{figure}
	\includegraphics[width=0.9\linewidth]{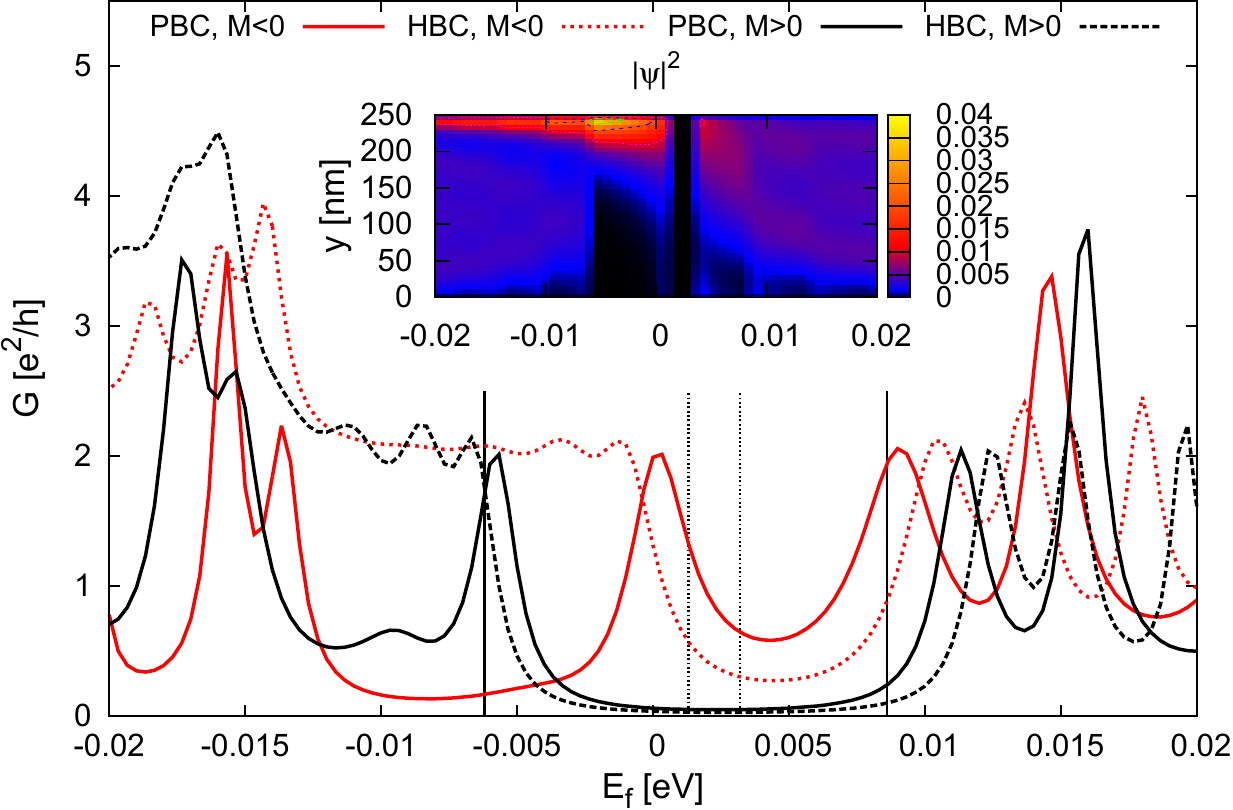}
	\caption{The conductance as a function of the Fermi energy in the central part of the junction. The Fermi energy in the leads is $C=-2.5\unit{eV}$. 
The sample is $250\unit{nm}$ wide and $100\unit{nm}$ long.  The vertical lines indicate the bulk gap (solid) and the mini-gap (dashed). 
The inset shows the density of states for spin-up electrons as a function of the Fermi energy and the $y$-coordinate in an infinite wire and for the inverted regime. 
The mini-gap can be identified by the lack of any propagating solution. 
 }
	\label{fig:Efplot}
\end{figure}

Another way to find differences between the topologically trivial and non-trivial regimes is by analyzing the conductance as a function of the Fermi energy for a sample of fixed length. For PBC the two regimes show the same behavior as long as the Fermi energy in the sample lies within conduction or valence band \cite{Novik10}. Only within the gap the QSHI can be distinguished from the normal insulator by a conductance peak around $E_f = 0$. For HBC we find the same behavior as within PBC  in regimes where the bulk states dominate the transport, i.e. in wide and short QSHI junctions. When we increase the length of the QSHI above the decay length of the evanescent modes, the conductance is governed by the edge state contribution. In this limit within PBC the signal goes to zero while for HBC the conductance is $2e^2/h$ in wide wires where Fabry-Perot oscillations are negligible. In narrow and short samples however the transport is mediated by a mixture of edge and bulk states. Exactly this  latter limit
is shown in Fig.~\ref{fig:Efplot} where the Fermi energy dependence of the conductance is presented. In Fig.~\ref{fig:Efplot} , the bulk gap is indicated by two black vertical lines. 
It is larger than $2 \vert M\vert$ due to quantum confinement. 
Additionally the dashed vertical lines present the energy range, in which the overlap of the edge states produces a gap in the system \cite{Zhou08}. We will refer to it as mini-gap from here on. One can notice that outside of the bulk gap (solid lines) the signals of the normal and the inverted regimes behave similar, like it was found for wide structures \cite{Novik10, Recher10}. Within the gap however and for the PBC,
 a clear conductance peak in the inverted regime around $E_f = 0$ is seen, while in the normal regime the transport is almost completely suppressed. For HBC one observes a similar behavior for the normal regime, but the existence of the edge states changes the signal drastically for $M<0$. Within the gap and for energies smaller than zero, we find the quantized conductance of the edge states. Around zero and for positive energies the signal is strongly reduced and resembles that of the bulk modes. This behavior can be understood by looking at the density of states in an infinite wire in the inverted regime. In the inset to Fig.~\ref{fig:Efplot} we show the corresponding $\vert\psi\vert^2$ as a function of the $y$-coordinate and the Fermi energy. As before we consider just one of the blocks, so that only one edge state (spin-up) is shown. The mini-gap as well as the bulk gap can be easily seen in this contour plot. We find that the edge state spreads over the whole sample shortly before the mini-gap opens. We have seen that the overlap of the edge state crucially influences the transport in metal/QSHI/metal junctions. Therefore let us note that the penetration depth of the edge state is in general a function of the energy\cite{Zhou08, Krueckl2011, Lu2011}, like it is shown in the inset of Fig.~\ref{fig:Efplot}. For $D$ finite and smaller than zero this behavior is asymmetric. For negative energies the edge state is strongly localized at the upper sample border. The small penetration depth leads to the fact that the edge state exists also at energies outside of the energy gap. For our parameters it merges to the valence band at energies of about $E_f= -80\unit{meV}$. 
 Up to this energy the strong localization ensures a good quantization which we observe in the conductance. In contrast, on the other side of the mini-gap ($E_F> 3\unit{meV}$) the edge state is  widely spread, so that a strong backscattering occurs. Moreover it merges quickly to the bulk states around $E_f = 10\unit{meV}$. Therefore we believe that at these energies the transport is dominated by evanescent bulk states. Hence in this regime the conductance resembles the shape of the PBC signal.\\

\section{Disorder}
\label{sect:disorder}
In this section we want to analyze the robustness of the conductance signal in the presence of scalar disorder. Disorder breaks the translational invariance of the sample which makes the application of the wave matching routine impossible. 
We go around this problem by using tight-binding calculations within the Landauer-B\"uttiker formalism \cite{Datta2007, Hankiewicz2004, Bruene2010}
In the tight-binding approach we discretize the sample using a finite grid of say $N_x\times N_y$ points while the hard wall boundary conditions are naturally implemented by missing bonds at the borders.
In principle we could have performed the whole analysis, which we reported in this paper, using tight-binding calculations. However, to check the validity of the conclusions for the experimentally relevant  setups
one needs to simulate large structures. Further for a good approximation within the tight-binding model one needs a small lattice constant $a_L$, i.e. a small distance between neighboring sites, because many fast decaying evanescent modes have to be taken into account. As a consequence we would have to use large systems with up to $(2000\times200\times2)^2$ (width $\times$ length $\times$ degrees of freedom) lattice points, which need extraordinary amount of memory and CPU time. \\

\begin{figure}
\includegraphics[width=0.9\linewidth]{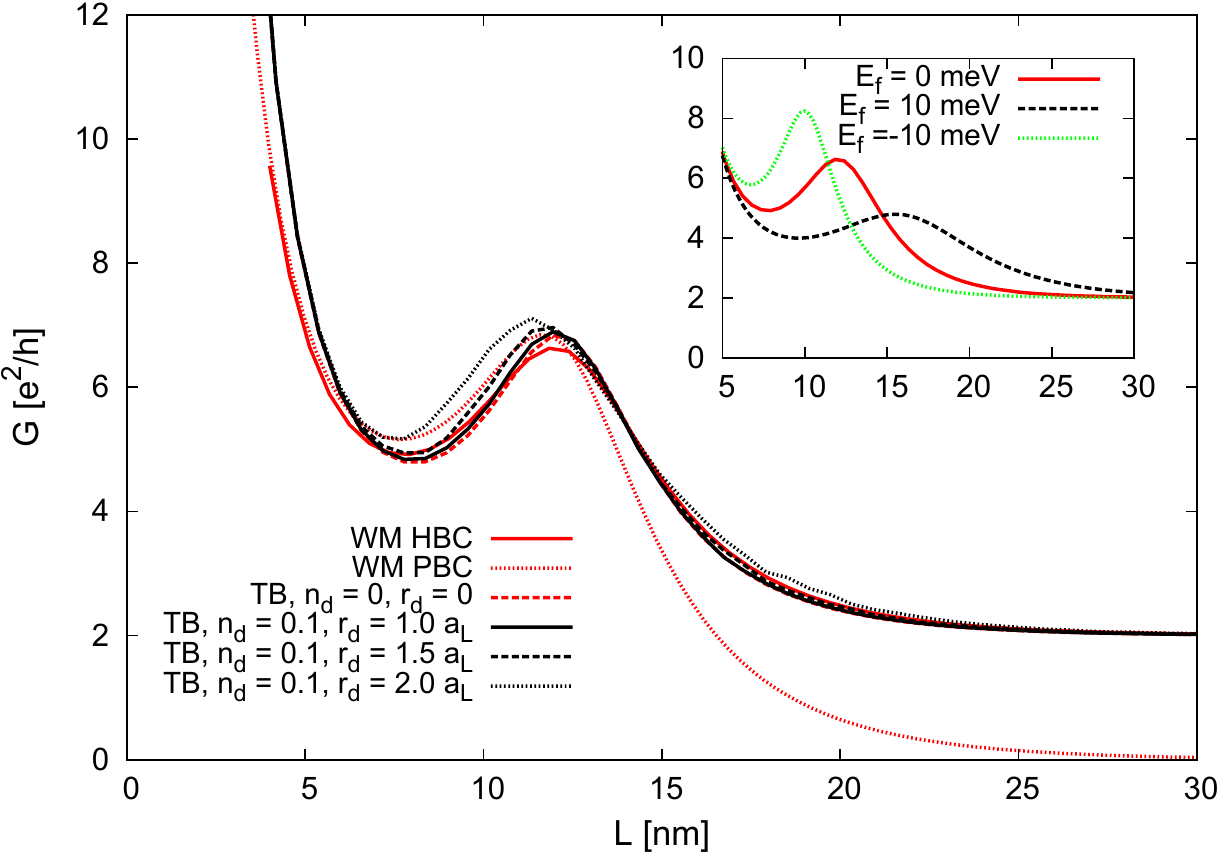}
\caption{The charge conductance as a function of the sample length. We compare the wave matching (WM) approach for HBC (solid  red/gray line) and for PBC (dotted red/gray line) with the tight-binding calculations (dashed red/gray line) and find good coincidence. Adding scalar disorder with Gaussian correlations influences the signal slightly but does not destroy the conductance peak. We used $W = 150 \unit{nm}$, $C= -2.5 \unit{eV}$ and $M = 31 \unit{meV}$. The disorder strength was set to $W_d = 0.06\unit{eV}$ for the black lines. The inset shows the influence of the Fermi energy $E_f$ inside the sample region.}
\label{fig:disorder}
\end{figure}

To analyze the influence of disorder let us here vary the sample structure and the parameters, similarly as it was performed for PBC in Ref.~\onlinecite{Recher10}.   
 First we decrease the width of the sample to $150\unit{nm}$. For PBC this reduction should not change the position of the maximum.
 However, for HBC the overlap of the edge state wave functions influences the signal  in narrow wires drastically. 
The overlap can  be reduced by increasing the bulk gap, i.e. the $M$ parameter as it was shown in Ref.~\onlinecite{Zhou08}.
 Further Eq.~\eqref{eq:Lmax} shows that a larger $M$ parameter leads to a smaller $L_{\text{max}}$.
Therefore, in the following by increasing $M$ by the factor of 10 to $M= -31.0 \unit{meV}$,  we expect the maximum around $10\unit{nm}$ instead of $100\unit{nm}$. 
Indeed Fig.~\ref{fig:disorder} shows the conductance maximum at about $12\unit{nm}$. 
First we compare the tight-binding (TB) calculations in the absence of disorder (red/gray dashed line in Fig.~\ref{fig:disorder}) with the signal computed in the wave matching (WM) formalism (red/gray solid line). 
The signals of the two models almost coincide when we choose $N_y = 250$ grid points. The number of points in $x$-direction varies with $L$, since we fix the lattice constant to $a_L = 150 \unit{nm}/251$. Therefore indeed the above described  procedure does not change the results quantitatively but shifts the position of the conductance maximum to system sizes which can be easily calculated within tight binding model.
Further the conductance for PBC (red/gray dotted line) coincide with the signal from tight-binding calculations  indicating that the overlap between the edge states is small.\\
Scalar disorder is included by choosing randomly $N_d = n_d N_x N_y$ ($n_d \in [0,1]$) lattice sites. For our analysis here we fixed  $n_d=0.1$.
 On each of the selected points we shift the energy randomly by $V_d \in [-W_d/2, W_d/2]$. The strongest disorder we can put without coupling the valence to the conduction band therefore is $W_d = 2 \vert M \vert$.
 The disorder is modeled by a Gaussian-like function with the decay length $r_d$ given in units of the lattice constant 
$a_L$. All signals including finite disorder have been averaged over 100 different disorder realization.\\
In Fig.~\ref{fig:disorder} we focus on the constant disorder strength $W_d = 0.06 \unit{eV}$  varying the decaying length of the Gaussian disorder from $r_d = 1 a_L$ (solid black line) to $2 a_L$ (dotted black line).
 For long sample lengths $L$, the disorder has no influence on the quantized conductance of the edge states, because the scalar disorder cannot couple the counter propagating edge channels. 
When evanescent waves contribute to the transport at smaller $L$ the disorder has a visible influence on the signal. 
The influence of the disorder increases with an increase of disorder decay length. It influences the signal by shifting and increasing the maximum of conductance. 
To understand this effect we need to remember that the transport for short QSHI lengths is dominated by evanescent modes. 
In this regime the extension of  impurities can be comparable with the sample length introducing large fluctuations of energies over a relatively large part of the sample which can cause a shift of the conductance maximum . 
Indeed the inset to Fig.~\ref{fig:disorder} shows the conductance signal shift to left or right depending on the Fermi energy inside the QSHI. 
Despite these small changes in the conductance behavior, the conductance signal is robust against disorder for HBC.\\

\section{Conclusion}
\label{sect:summary}
We analyzed the interplay of edge and bulk states in  metal/quantum spin-Hall insulator (QSHI)/metal junctions as a function of the size of QSHI part as well as the Fermi energy.
 We found that for short and wide QSHI part, the edge states are not populated due to conductance mismatch between 
highly doped leads and QSHI regime, and the transport occurs through the evanescent modes.
In this regime, the topologically non-trivial insulator can be distinguished from the trivial insulator by the appearance of the  distinct conductance maximum at small QSHI length 
and the conductance  peak  in the gap when the Fermi energy changes for the constant length.
 Moreover, we showed that the origin of this conductance maximum is surprisingly the formation of an effectively propagating state with the real $k_x$ vector built from evanescent modes. 

In contrast, in the wide and long samples the transport is driven by spin edge channels while  in the  intermediate size regime the conductance signal is built from both bulk and edge states 
in the ratio which depends on the width of the sample. In this latter regime the combination of the Fabry-Perot oscillations (coming from the overlap of edge states) and bulk modes  
changes the position and the shape of the maximum. 

Our predictions are  robust against scalar disorder. Therefore  the measurements of  systems with diferent geometrical aspect ratios should allow to distinguish
 the bulk and edge state contributions to the transport.   Further the detection of the  bulk conductance peak as a  function of Fermi energy or the length for short and wide QSHI junctions, 
should be an alternative method to distinguish between topologically trivial and non-trivial insulators in HgTe QWs as well as Bi$_2$Se$_3$ thin films.

\textbf{Acknowledgements}\\ We acknowledge interesting discussions with B. Trauzettel, P. Recher, E. Novik, D.G. Rothe and P. Michetti. 
This work was funded through DFG Grant HA5893/1-2. We thank Leibniz Rechenzentrum Munich for providing computing resources. 


\end{document}